\documentclass[fp,dvipdfmx]{jpsj3}
\renewcommand{\a}{\alpha}

\newcommand{\bea}{\begin{eqnarray}}
\newcommand{\eea}{\end{eqnarray}}
\newcommand{\f}[2]{\frac{#1}{#2}}
\newcommand{\eq}{&=&}
\newcommand{\nn}{\nonumber \\ }
\newcommand{\ve}{\varepsilon}

\newcommand{\area}{\int_{-\infty}^\infty }

\renewcommand{\l}{\lambda}
\newcommand{\p}{\partial}
\newcommand{\pp}[2]{\f{\p #1}{\p #2}}

\renewcommand{\H}{{\cal H}}
\renewcommand{\L}{{\cal L}}
\allowdisplaybreaks[1]
\usepackage{color}

\renewcommand{\l}{\lambda}

\newcommand{\habu}[1]{}
\allowdisplaybreaks[1]

\title{
Random matrix approach for primal-dual portfolio optimization problems
}

\author{
{Daichi Tada${}^{\dagger}$, 
Hisashi Yamamoto${}^{\dagger}$ and 
Takashi Shinzato\thanks{Corresponding author, shinzato@eng.tamagawa.ac.jp}
}}
\inst{
${}^{\dagger}$
Graduate School of System Design, Tokyo Metropolitan University, \\
Hino, Tokyo 191-0065, Japan\\
${}^{\ast}$
Department of Management Science, College of Engineering, 
Tamagawa University, \\ Machida, Tokyo 194-8610, Japan\\
\smallskip
{\rm (Received September 14, 2017; accepted *** 1, 2017)}
} 

\abst{
In this paper, we revisit the portfolio optimization problems of the minimization/maximization of investment risk under constraints of budget and investment concentration (primal problem) and the maximization/minimization of investment concentration under constraints of budget and investment risk (dual problem) for the case that the variances of the return rates of the assets are identical.
We analyze both optimization problems by using the Lagrange multiplier method and the random matrix approach. 
Thereafter, we compare the results obtained from our proposed approach with the results obtained in previous work.
Moreover, we use numerical experiments to validate the results obtained from the replica approach and the random matrix approach as methods for analyzing both the primal and dual portfolio optimization problems.
}

\kword{mean variance model, primal and dual problems, investment risk, investment concentration, random matrix approach, Stieltjes transformation, replica analysis}

\begin{document}
\maketitle

\section{Introduction}
When investors invest in financial instruments traded in the securities market, it is important for them to implement an appropriate risk management strategy since the return of an asset is uncertain.
For this reason, Markowitz formulated the portfolio optimization problem to mathematically analyze risk management based on diversified investment \cite{1}.
Following his pioneering work, many mathematical models and analytical approaches have been reported in operations research \cite{2,3}. 
Recently, the properties of the optimal portfolio of the portfolio optimization problem have been actively explored using the analytical approaches developed in the cross-disciplinary research fields involving econophysics and statistical mechanical informatics.
For instance, Laloux {\it et al.} investigated the statistical structure of the empirical correlation matrix for 406 assets in the ${\rm S \& P500}$ based on daily normalized returns for a total of 1309 days during 1991--1996 and proposed a technique for pruning the noise from the empirical correlation matrix by comparing the eigenvalue distribution of the empirical correlation matrix with that of a random matrix \cite{4,5}. 
Plerou {\it et al.} analyzed the eigenvalue spacing distribution to investigate whether the empirical correlation matrix exhibits the universal statistical properties predicted by random matrix theory. 
For that purpose, they used stock price data of U.S. publicly traded companies over the 2-year period 1994-1995 \cite{6}. 
Pafka {\it et al.} quantitatively evaluated the correlation between assets by comparing the eigenvalue distribution of the empirical correlation matrix of return calculated from actual data with  the Mar$\check{{\rm c}}$enko-Pastur distribution \cite{7}.
Ciliberti {\it et al.} applied the replica analysis method to the mean-absolute deviation model and the expected shortfall model, and analyzed the phase transition that occurs when the number of data periods is large relative to the number of assets \cite{8}.
Shinzato showed that the minimal investment risk and its investment concentration satisfy the self-averaging property by using the large deviation principle, and he compared the minimal investment risk per asset derived using replica analysis with the minimal expected investment risk per asset derived using operations research and concluded that a portfolio which can minimize the expected investment risk does not necessarily minimize the investment risk \cite{9}.
Additionally, Shinzato analyzed the minimization of investment risk under constraints of budget and investment concentration along with the dual problem by using replica analysis in order to clarify their primal-dual structure \cite{10,11}.
Varga-Haszonits {\it et al.} examined the minimization of risk function defined by the sample variance under constraints of the budget and the expected return by using replica analysis and analyzed the stability of the replica symmetric solution \cite{12}.
Shinzato examined the optimal portfolio which can minimize the investment risk with short selling using replica analysis and disclosed the phase transition of this investment system \cite{shinzato-short-selling}.
Kondor {\it et al.} also analyzed the minimization of investment risk under constraints of budget and short selling in the case that the variances of the return rates of the assets are not identical by using replica analysis and reconfirmed that this disordered system involves a phase transition \cite{13}.

Although Ciliberti {\it et al.}, Kondor {\it et al.}, and Shinzato {\it et al.} have all studied the properties of the optimal solution of the portfolio optimization problem by using replica analysis, it is generally known that the validities of several methods used in replica analysis, for example, the {analytic continuation} of the replica number, have not been guaranteed theoretically\cite{14,shinzato-arxiv}.
Therefore, we need to validate results obtained from replica analysis by another approach that is mathematically guaranteed, such as the random matrix approach. 
In addition, as mentioned above, the application of random matrix theory to the portfolio optimization problem is mainly based on the evaluation of the correlation matrix of the return rate, whereas its application to investment risk and investment concentration of the optimal portfolio has not been sufficiently examined.

In the present paper, we reassess the minimal/maximal investment risk per asset with budget and investment concentration constraints and the maximal/minimal investment concentration
with constraints of budget and investment risk using the random matrix approach.
Moreover, we consider whether it can be applied to the evaluation of the minimal/maximal investment risk per asset and investment concentration derived by the asymptotic eigenvalue distribution when the number of assets is sufficiently large but finite (not the thermodynamical limit).

This paper is organized as follows:
In the next section, we formulate the minimization/maximization of investment risk under constraints of budget and investment concentration (primal problem) and the counterpart problems, the maximization/minimization of investment concentration under constraints of budget and investment risk (dual problem).
In Sec.\ III, we analyze these two problems using the Lagrange multiplier method and the random matrix approach.
In Sec.\ IV, numerical experiments for the portfolio optimization problems  are carried out  to validate the replica approach and the random matrix approach.
Finally, in Sec.~V, we devote our conclusions and future work.
\section{Model Setting}
The present study considers the optimal diversification investment based on the mean variance model as the primal problem and dual problem with $N$ assets over $p$ periods in a stable investment market where short selling is not regulated.
The position on asset $i\ (=1,2,\cdots,N)$ is denoted as $w_{i}$, and the portfolio of $N$ assets is represented as $\vec{w}=(w_1,w_2,...,w_N)^{\rm {\rm T}} \in {\bf R}^{\rm N}$, where notation ${\rm T}$ indicates the transpose operator. 
Note that $w_i$ takes any real number due to there being no restriction on short selling.
Furthermore, $\bar{x}_{i \mu}$ shows the return rate of asset $i$ at period $\mu (=1,2,\cdots,p)$, where the return rates are independent and identically distributed with mean $E[\bar{x}_{i \mu}]$ and variance $V[\bar{x}_{i \mu}]=1$.

Herein, the objective function of the primal problem, $\H_{P}(\vec{w})$, is defined as follows:
\begin{eqnarray}
\label{eq1}
\H_{P}(\vec{w})\eq
\frac{1}{2N}\sum_{\mu=1}^p \left(\sum_{i=1}^N \bar{x}_{i \mu}w_i-\sum_{i=1}^NE[\bar{x}_{i \mu}]
w_i \right)^2,
\end{eqnarray}
where $\sum_{i=1}^N \bar{x}_{i \mu}w_i$ is the return rate of the portfolio at period $\mu$ and $\sum_{i=1}^N E[\bar{x}_{i \mu}]w_i$ is its expectation.
We will call $\H_{P}(\vec{w})$ in Eq.~(\ref{eq1}) investment risk hereafter.
Investment risk can be rewritten as follows:
\begin{eqnarray}
\label{eq4}
\H_{P}(\vec{w})\eq\frac{1}{2}\vec{w}^{{\rm T}}J\vec{w},
\end{eqnarray}
where $J=\{J_{ij}\} \in {\bf R}^{N \times N}$ is the variance-covariance matrix (or Wishart matrix) with components $i,j$ given by $J_{ij}=\frac{1}{N}\sum_{\mu=1}^p x_{i \mu} x_{j \mu}$ in terms of the modified return rate $x_{i \mu}=\bar{x}_{i \mu}-E[\bar{x}_{i \mu}]$.
In the primal problem, the portfolio $\vec{w}$ is under the constraints defined as follows:
\begin{eqnarray}
\label{eq5}
\sum_{i=1}^N w_i&=&N, \\
\label{eq7}
\sum_{i=1}^N w_i^2&=&N \tau \qquad \tau \ge 1,
\end{eqnarray}
where Eq.~(\ref{eq5}) is the budget constraint and Eq.~(\ref{eq7}) is the investment concentration constraint.
$\tau$ in Eq.~(\ref{eq7}) is a constant which characterizes an investment concentration of risk management strategy.
Investment concentration $q_w=\f{1}{N}\sum_{i=1}^Nw_i^2$  is an index for assessing the dispersion of the portfolio $\vec{w}$, and it shows the achievement of diversified investment in the same way as the Herfindahl-Hirschman index \cite{11}.
In addition, the budget constraint in Eq.~(\ref{eq5}), which is different from the budget constraint $\sum_{i=1}^N w_i=1$ used in operations research, is rescaled so that $w_{i}$ is $O(1)$ in the limit as the number of assets $N$ approaches infinity, and  it is possible to give a statistical interpretation to investment concentration $q_w=\f{1}{N}\sum_{i=1}^Nw_i^2$ by using Eq.~(\ref{eq5}) \cite{11}.
Moreover, the subset of feasible portfolios, that is, those satisfying Eqs.~(\ref{eq5}) and (\ref{eq7}), ${\cal W}_{P}$, is defined as follows:
\begin{eqnarray}
\label{eq2}
{\cal W}_{P}\eq
\left\{\vec{w} \in {\bf R}^{N}
\left|
\vec{w}^{\rm T}\vec{e} =N,\ \vec{w}^T \vec{w}=N \tau 
\right.
\right\},
\end{eqnarray}
where $\vec{e}=(1,1,\cdots,1)^{\rm T} \in {\bf R}^{\rm N}$ is the ones vector.
For the primal problem, we consider the minimal investment risk per asset $\varepsilon_{\min}$ and the maximal investment risk per asset $\varepsilon_{\max}$:
\begin{eqnarray}
\label{eq6}
\varepsilon_{\min}\eq\lim_{N \rightarrow \infty} \min_{\vec{w} \in {\cal W}_{P}} \left \{\frac{1}{N}\H_{P}(\vec{w}) \right \},\\
\label{eq72}
\varepsilon_{\max}\eq\lim_{N \rightarrow \infty} \max_{\vec{w} \in {\cal W}_{P}} \left \{\frac{1}{N}\H_{P}(\vec{w}) \right \}.
\end{eqnarray}

Note, the primal problem and the dual problem have  been evaluated already by replica analysis in previous work and the validity of the findings by replica analysis were verified by numerical experiments \cite{11}.
However, the validity of the analytical approach of replica analysis has not been mathematically guaranteed.
Thus, it is expected that there may be some resistance to conducting investment actions based on the results of replica analysis. 
For this reason, we reevaluate the primal problem and the dual problem by using the random matrix approach guaranteed mathematically.
In previous work, using replica analysis, the minimal investment risk per asset $\varepsilon_{\min}$ was derived as follows:
\begin{eqnarray}
\label{eq44}
\varepsilon_{\min}&=&\left\{ \begin{array}{ll} 
\frac {\alpha \tau+\tau-1-2\sqrt {\alpha \tau \left( \tau -1\right) }}{2} & 1-\frac{1}{\tau} \leq \alpha  \\ \\
0 & {\rm otherwise},
\end{array}\right.
\end{eqnarray}
where the period ratio $\alpha=p/N \sim O(1)$ is used \cite{10}. 
In addition, the maximal investment risk per asset $\varepsilon_{\max}$ is as follows:
\begin{eqnarray}
\label{eq43}
\varepsilon_{\max}\eq\frac {\alpha \tau+\tau-1+2\sqrt {\alpha \tau \left( \tau -1\right) }}{2}\qquad\alpha>0.
\end{eqnarray}

Similarly, we set the dual problem corresponding to this primal problem.
In this case, the objective function, $\H_{D}(\vec{w})$, is defined as follows:
\begin{eqnarray}
\label{eq60}
\H_{D}(\vec{w})\eq\frac{1}{2} \sum_{i=1}^N w_{i}^{2}.
\end{eqnarray}
This function corresponds to the investment concentration in Eq.~(\ref{eq7}), which is one of the constraints of the primal problem.
The constraints of the dual problem are the budget constraint in Eq.~(\ref{eq5}) and the investment risk constraint defined as follows:
\begin{eqnarray}
\label{eq61}
\frac{1}{2}\vec{w}^{{\rm T}}J\vec{w}\eq
N \kappa \varepsilon_{0}.
\end{eqnarray}
This equation uses the minimum investment risk per asset in the portfolio optimization problem with only the budget constraint imposed, $\ve_{0}=\f{\a-1}{2}$ \cite{9}.
This constraint implies that the investment risk for $N$ assets is $\kappa (\ge 1)$ times the minimal investment risk imposed on the budget for $N$ assets, which is $N \ve_{0}$. 
We call $\kappa$ the risk coefficient.
Then the feasible portfolio subset satisfying Eq.~(\ref{eq5}) and Eq.~(\ref{eq61}) is defined as follows:
\begin{eqnarray}
\label{eq70}
{\cal W}_{D}\eq\left \{\vec{w} \in {\bf R}^{N}
\left|
\vec{w}^{\rm T} \vec{e}=N,\ \frac{1}{2}\vec{w}^{{\rm T}}J\vec{w}=N \kappa \varepsilon_{0}
\right.
\right\}. \quad
\end{eqnarray}
For the dual problem, we consider the maximal investment concentration $q_{w, \max}$ and the minimal investment concentration $q_{w, \min}$:
\begin{eqnarray}
\label{eq50}
q_{w,\max}\eq\lim_{N \rightarrow \infty} \max_{\vec{w} \in {\cal W}_{D}} \left \{ \frac{2}{N}\H_{D}(\vec{w}) \right \},\\
\label{eq71}
q_{w,\min}\eq\lim_{N \rightarrow \infty} \min_{\vec{w} \in {\cal W}_{D}} \left \{\frac{2}{N}\H_{D}(\vec{w}) \right \}.
\end{eqnarray}
In previous work\cite{11}, using replica analysis, the maximal investment concentration $q_{w, \max}$ was also analyzed as 
\begin{eqnarray}
\label{eq47}
q_{w,\max}\eq\frac{\left( \sqrt{\alpha \kappa}+\sqrt{\kappa-1} \right)^2}{\alpha-1}\qquad\alpha>1,
\end{eqnarray}
and the minimal investment concentration $q_{w, \min}$ was assessed as 
\begin{eqnarray}
\label{eq48}
q_{w,\min}\eq\frac{\left( \sqrt{\alpha \kappa}-\sqrt{\kappa-1} \right)^2}{\alpha-1}\qquad\alpha>1.
\end{eqnarray}

In the replica analysis of previous work\cite{11}, for instance, in the case of the primal problem, the minimum investment risk per asset $\varepsilon_{\min}$ was evaluated using the following equation:
\begin{eqnarray}
\label{eq90}
\varepsilon_{\min}\eq\lim _{\beta \rightarrow \infty} \left \{ -\frac{\partial}{\partial \beta} \lim _{N \rightarrow \infty} \frac{1}{N} E[\log Z] \right \},
\end{eqnarray}
where the partition function $Z$ is defined using the Boltzmann distribution of the inverse temperature $\beta$ as
\begin{eqnarray}
\label{eq91}
Z\eq\int_{\vec{w} \in {\cal W}_{P}} d\vec{w} e^{-\beta \H_p(\vec{w})},
\end{eqnarray}
and the notation $E[\cdot]$ means the expectation with respect to the modified return rate $ x_ {i \mu} $, which is called the configuration average.
In the process of evaluating $E[\log Z]$ included in the right-hand side of Eq.~(\ref{eq90}), an identity called a replica trick,
\begin{eqnarray}
\label{eq49}
E[\log Z]\eq\lim _{n\rightarrow 0} \frac{\partial}{\partial n}\log E[Z^n],
\end{eqnarray} 
is often used. 
In replica analysis, one first assumes that the replica number in Eq.~(\ref{eq49}), $n$, is an integer and that one can implement a configuration average of $Z^n$ (that is, calculate $E[Z^n]$). 
Then, we assume that the replica number is real and that {analytic continuation 
of $E[Z^n]$
can be executed.} 
However, the validity of the analytic continuation with respect to replica number from an integer to a real number has not yet been mathematically guaranteed for portfolio optimization problems, which is also the case for many problems in cross-disciplinary research fields  \cite{14}.
Therefore, we reexamine the primal problem and the dual problem mathematically in the present work by using the Lagrange multiplier method and the random matrix approach, as alternatives to replica analysis.
\section{Lagrange multiplier method and random matrix approach}
In this section, we consider the primal problem and the dual problem via the Lagrange multiplier method and the random matrix approach \cite{Boyd,Nocedal,20}. 

\subsection{\label{sec3.1}Minimal investment risk with budget and investment concentration constraints}
First, we examine the minimal investment risk with budget and investment concentration constraints as the primal problem. 
The Lagrange function corresponding to the minimization problem of the investment risk $\L_{P}(\vec{w},k,\theta)$ is defined using Lagrange multipliers $k,\theta $ as follows:
\begin{eqnarray}
\label{eq12}
\L_{P}(\vec{w},k,\theta)\eq
\H_{P}(\vec{w})+k(N-\vec{w}^{\rm T}\vec{e})+\frac{\theta}{2}(N\tau-\vec{w}^{\rm T}\vec{w}), \nonumber \\
{\rm s.t.}&&k \in {\bf R} \nonumber \\
&&\theta \leq \lambda_{i}\qquad \forall i=1,2,...,N,
\end{eqnarray}
where $\lambda_{i}$ represents the eigenvalue of the Wishart matrix $J$ and 
the constraint in Eq.~(\ref{eq12}) is a condition to guarantee that the Lagrange function $\L_{P}(\vec{w},k,\theta)$ is convex with respect to $\vec{w}$ \cite{Boyd,Nocedal,20}.
This constraint can be rewritten as $\theta \leq \lambda_{\min}=\min_{1 \leq i \leq N}\lambda_{i}$, for which $\lambda_{\min}$ in the limit as $N$ approaches infinity is known:  
\begin{eqnarray}
\label{eq51}
\lambda_{\min}&=&\left\{ \begin{array}{ll} 
(1-\sqrt{\alpha})^2 & \alpha \geq 1 \\
0 & 0<\a<1{,}
\end{array} \right. 
\end{eqnarray}
where the asymptotical spectrum of $J$ (Mar$\check{{\rm c}}$enko-Pastur law) is used in the limit as $N$ goes infinity keeping $\a=p/N \sim O(1)$ \cite{15}.
Note that the Lagrange multiplier method can also handle the investment risk with only the budget constraint imposed when $\theta=0$ (see Appendix \ref{app-c} for details).
From the extremum equations $\frac{\partial \L_{P}(\vec{w},k,\theta)}{\partial \vec{w}}=0$ and $\frac{\partial \L_{P}(\vec{w},k,\theta)}{\partial k}=0$,  
\begin{eqnarray}
\label{eq13}
\vec{w}^{\ast}&=&k^{\ast}(J-\theta I_N)^{-1}\vec{e}, \\
\label{eq14}
k^{\ast}&=&\frac{1}{S_N(\theta)},
\end{eqnarray}
are obtained, where $I_N$ in Eq.~(\ref{eq13}) is the $N \times N$ identity matrix, $J-\theta I_N$ is a regular matrix because of Eq.~(\ref{eq12}), and $S_N(\theta)$ is defined as follows:
\begin{eqnarray}
\label{eq16}
S_N(\theta)\eq
\frac{\vec{e}^{\rm T}(J-\theta I_N)^{-1}\vec{e}}{N}.
\end{eqnarray}
Let us employ the properties of $S_N(\theta)$ in the limit as $N$ approaches infinity in order to analyze briefly the optimal $\theta$ of the Lagrange function.
The singular value decomposition of the return rate matrix $X= \left\{\frac{x_{i \mu}}{\sqrt{N}} \right\} \in {\bf R}^{N\times p}$ is expressed as $X=UDV^{\rm T}$ using $N \times N$ orthogonal matrix $U$, $p \times p$ orthogonal matrix $V$, and diagonal matrix $D={\rm diag} \{d_{i} \} \in {\bf R}^{N\times p}$, where $d_{i}=\pm\sqrt{\lambda_{i}}$ are the singular values. Then the Wishart matrix $J$ is rewritten as $J={XX}^{\rm T}=UDD^{\rm T}U^{\rm T}$.
From this, $S_N(\theta)$ can be rewritten as follows:
\begin{eqnarray}
S_N(\theta)&=&\frac{1}{N}\sum_{k=1}^{N} \frac{u_k^2}{\lambda_k-\theta} \nonumber \\
&=&\int_{- \infty}^{\infty} \frac{1}{\lambda - \theta} \frac{1}{N} \sum_{k=1}^N u_k^2  \delta(\lambda - \lambda_k)d\lambda,
\end{eqnarray}
where $\vec{u}$ is defined as $\vec{u}=U^{\rm T}\vec{e}=(u_1,u_2,...,u_N)^{\rm T} \in {\bf R}^{N}$ and the Dirac delta function $\delta(x)$ is used.
Since $\vec{u}$ satisfies $\vec{u}^{\rm T}\vec{u}=\vec{e}^{\rm T}\vec{e}=N$, it is known that $u_k$ is asymptotically, independently, and identically  distributed according to the standard normal distribution 
in the limit that $N$ goes to infinity \cite{16,17}. 
Further, $S_N(\theta)$ has the self-averaging property in the thermodynamic limit of $N$, so that $S_N(\theta)$ converges to $S(\theta)=\lim_{N \to \infty}E[S_N(\theta)]$, which is given by the following equation:
\begin{eqnarray}
\label{eq18}
S(\theta)&=&\int_{- \infty}^{\infty} \frac{\rho(\lambda)}{\lambda - \theta} d\lambda,
\end{eqnarray}
where $\rho(\lambda)\ (=\lim_{N \to \infty}\frac{1}{N} \sum_{k=1}^N \delta(\lambda - \lambda_k))$ is an asymptotic eigenvalue distribution of the Wishart matrix $J$, that is, the Mar$\check{{\rm c}}$enko-Pastur law \cite{15}.
The integral defined in Eq.~(\ref{eq18}), $S(\theta)$, is generally called the Stieltjes transform for the Mar$\check{\rm c}$enko-Pastur law \cite{18,19}.
For the details of the evaluation of Eq.~(\ref{eq18}), see Appendices A and B.

Based on the above argument, the investment risk per asset $\ve(\theta)$ can be represented as a function of $\theta$ in the thermodynamic limit of $N$ as follows:
\begin{eqnarray}
\label{eq27-1}
\ve(\theta)&=&\lim_{N\to\infty}\f{1}{N}\L_{P}(\vec{w}^*,k^*,\theta) \nonumber \\
&=&\frac {1}{2} \left( \frac {1}{S \left( \theta \right)}+\tau \theta \right).
\end{eqnarray}
Then the minimal investment risk per asset $\varepsilon_{\min}$ is given by the following extremum:
\begin{eqnarray}
\label{eq20}
\varepsilon_{\min}&=&\sup_{\theta \leq \lambda_{\min}}
\ve(\theta).
\end{eqnarray}
From this, it is possible to analytically evaluate $\ve_{\min}$ by using Eq.~(\ref{eq27-1}) and Eq.~(\ref{A5}).
{Notice that this equation is equivalent to the minimization of investment risk per asset with respect to portfolio $\vec{w}$.}
In the following, we consider the minimal investment risk per asset $\varepsilon_{\min}$ in the ranges of (i) $\alpha \ge 1$ and (ii) $0 < \alpha < 1$.

\begin{description}
\item[(i)]
First, in the range $\alpha \ge 1$, 
$\lambda_{\min}=(1-\sqrt{\alpha})^2$ is given in Eq.~(\ref{eq51}) and there exists $\theta$ in the range $\theta \leq \lambda_{\min}$ such that $\frac{\partial \varepsilon}{\partial \theta}=0$.
This implies
\begin{eqnarray}
\label{eq27}
\theta^{\ast}&=&1+\alpha - \left(2\tau-1 \right) \sqrt{\frac{\alpha}{\tau(\tau-1)}}.
\end{eqnarray}
From this, the minimal investment risk per asset $\varepsilon_{\min}=\ve(\theta^*)$ is obtained as follows:
\begin{eqnarray}
\label{eq54}
\varepsilon_{\min}&=&\frac{\alpha \tau+\tau-1-2\sqrt {\alpha \tau \left( \tau -1\right) }}{2}.
\end{eqnarray}

\item[(ii)]
Next, in the range $0 < \alpha < 1$, 
$\lambda_{\min}=0$ is already given in Eq.~(\ref{eq51}). Then $\lim_{\theta \to \lambda_{\min}}\frac{\partial \varepsilon}{\partial \theta}=\frac{1}{2} \left( \tau-\frac{1}{1-\alpha} \right)$ is calculated.
When $\lim_{\theta \to \lambda_{\min}}\frac{\partial \varepsilon}{\partial \theta}<0$, that is, $1-\frac{1}{\tau}<\alpha<1$, since the investment risk per asset $\varepsilon(\theta)$ is maximal at $\theta^{\ast}$ given by Eq.~(\ref{eq27}), the same result as Eq.~(\ref{eq54}) is derived.
On the other hand, when $\lim_{\theta \to \lambda_{\min}}\frac{\partial \varepsilon}{\partial \theta}>0$, that is, $0<\alpha<1-\frac{1}{\tau}$, since 
the investment risk per asset $\varepsilon(\theta)$ is maximal at $\theta^{\ast}=\lambda_{\min}=0$, the minimal investment risk per asset $\varepsilon_{\min}=\ve(\l_{\min})$ is
\begin{eqnarray}
\label{eq55}
\varepsilon_{\min}=0.
\end{eqnarray}
\end{description}
By the above argument, 
$\theta^{\ast}=\arg \sup_{\theta \leq \lambda_{\min}} \varepsilon(\theta)$ is given by the following equation:
\begin{eqnarray}
\label{eq53}
\theta^{\ast}&=&\left\{ \begin{array}{ll} 
1+\alpha - \left(2\tau-1 \right) \sqrt{\frac{\alpha}{\tau(\tau-1)}} &1-\frac{1}{\tau}<\alpha  \\ \\
0  & {\rm otherwise}.
\end{array} \right.
\end{eqnarray}
Thus, the minimal investment risk per asset $\varepsilon_{\min}=\ve(\theta^*)$ is 
\begin{eqnarray}
\label{eq70}
\varepsilon_{\min} &=&\left\{ \begin{array}{ll} 
\frac {\alpha \tau+\tau-1-2\sqrt {\alpha \tau \left( \tau -1\right) }}{2} & 1-\frac{1}{\tau}<\alpha  \\ \\
0 & {\rm otherwise}.
\end{array} \right.
\end{eqnarray}
This result is consistent with the findings in previous work using replica analysis \cite{10}.

\subsection{\label{sec3.1-M}Maximal investment risk with budget and investment concentration constraints}
Next, we consider the maximal investment risk with budget and investment concentration constraints as the primal problem. 
The Lagrange function corresponding to the maximization problem of the investment risk $\L_{P}(\vec{w},k,\theta)$ is defined as follows:
\begin{eqnarray}
\label{eq71}
\L_{P}(\vec{w},k,\theta)\eq
\H_{P}(\vec{w})+k(N-\vec{w}^{\rm T}\vec{e})+\frac{\theta}{2}(N\tau-\vec{w}^{\rm T}\vec{w}) \nonumber \\
{\rm s.t.}&&k \in {\bf R} \nonumber \\
&&\theta \ge \lambda_{i}\qquad\forall i=1,2,...,N,
\end{eqnarray}
where the constraint in Eq.~(\ref{eq71}) is a condition to guarantee that the Lagrange function $\L_{P}(\vec{w},k,\theta)$ is concave with respect to $\vec{w}$ \cite{Boyd,Nocedal,20}.
This constraint can be rewritten as $\theta \ge \lambda_{\max}=\max_{1 \leq i \leq N}\lambda_{i}$, where $\lambda_{\max}$ in the thermodynamical limit of $N$ is known to be given by  
\begin{eqnarray}
\label{eq72}
\lambda_{\max}\eq(1+\sqrt{\alpha})^2,
\end{eqnarray}
which again uses the Mar$\check{\rm c}$enko-Pastur law \cite{15}.
Since the optimality of $\vec{w}$ and $k$ are already shown in the previous subsection (Eqs. (\ref{eq13}) and (\ref{eq14})), the maximal investment risk per asset $\varepsilon_{\max}$ is derived as follows:
\begin{eqnarray}
\label{eq73}
\varepsilon_{\max}&=&\inf_{\theta \ge \lambda_{\max}} \ve(\theta){,}
\end{eqnarray}
where $\ve(\theta)$ is in Eq. (\ref{eq27-1}). 
{Note that this equation is equivalent to the maximization of investment risk per asset with respect to portfolio $\vec{w}$.}
For any $\a>0$, since there exists $\theta^{\ast}$ of $\frac{\partial \varepsilon}{\partial \theta}=0$ in the range $\theta^{\ast} \ge \lambda_{\max}$, 
\begin{eqnarray}
\label{eq29}
\theta^{\ast}\eq 1+\alpha + \left(2\tau-1 \right) \sqrt{\frac{\alpha}{\tau(\tau-1)}},
\end{eqnarray}
and thus the maximal investment risk per asset $\varepsilon_{\max}=\ve(\theta^*)$ is as follows:
\begin{eqnarray}
\label{eq30}
\varepsilon_{\max}
&=&\frac {\alpha \tau+\tau-1+2\sqrt {\alpha \tau \left( \tau -1\right)}}{2}.
\end{eqnarray}
This derived result also is consistent with findings in previous work \cite{10}.
\subsection{Maximal investment concentration with constraints of budget and investment risk}
Next, we consider the maximal investment concentration with constraints of budget and investment risk as the dual problem. 
The Lagrange function corresponding to the maximization problem of the investment concentration $\L_{D}(\vec{w},h,\varphi)$ is defined using Lagrange multipliers $h,\varphi$ as follows: 
\begin{eqnarray}
\label{eq55}
\L_{D}(\vec{w},h,\varphi)\eq
\H_{D}(\vec{w})+\frac{h}{\varphi}
\left(\vec{w}^{\rm T}\vec{e}-N \right)
+\frac{1}{\varphi} \left(\kappa N \varepsilon_{0}-\frac{1}{2}\vec{w}^{{\rm T}}J\vec{w} \right),
\nonumber \\
{\rm s.t.}&&h\in {\bf R} \nonumber \\
&&\varphi^{-1}\lambda_{i} \ge 1\qquad\forall i=1,2,...,N,
\end{eqnarray}
where the constraint in Eq.~(\ref{eq55}) is a condition to guarantee that the Lagrange function $\L_{D}(\vec{w},h,\varphi)$ is concave with respect to $\vec{w}$ \cite{Boyd,Nocedal,20}.
This constraint can be rewritten as $0 \leq\varphi \leq \lambda_{\min}=(1-\sqrt{\alpha})^2$ by using the Mar$\check{{\rm c}}$enko-Pastur law in the limit that $N$ goes to infinity (we consider the maximal/minimal investment concentration in the range $\alpha \ge 1$ because $\varepsilon_0 \ge 0$).
From the extremum conditions $\frac{\partial \L_{D}(\vec{w},h,\varphi)}{\partial \vec{w}}=0$ and $\frac{\partial \L_{D}(\vec{w},h,\varphi)}{\partial h}=0$,
\begin{eqnarray}
\label{eq56}
\vec{w}^{\ast}&=&h^{\ast}(J-\varphi I_N)^{-1}\vec{e}, \\
\label{eq57}
h^{\ast}&=&\frac{1}{S_{N}(\varphi)},
\end{eqnarray}
are obtained and the investment concentration $q_{w}(\varphi)$ can be represented as a function of $\varphi$:
\begin{eqnarray}
\label{eq59}
\label{eq42}
q_{w}(\varphi)&=&\lim_{N \to \infty}\frac{2}{N}\L_{D}(\vec{w}^{\ast},h^{\ast},\varphi) \nonumber \\
&=&-\frac{1}{\varphi}\left( \frac{1}{S(\varphi)}-2\kappa \varepsilon_{0} \right).
\end{eqnarray}
Thus, the maximal investment concentration is derived from the following extremum:
\begin{eqnarray}
\label{eq74}
q_{w,\max}&=&\inf_{0 \leq\varphi \leq \lambda_{\min}}q_w(\varphi).
\end{eqnarray}
{Note that this equation is equivalent to the maximization of investment concentration with respect to portfolio $\vec{w}$.}
For any $\a>1$, since there exists $\varphi^{\ast}$ of $\frac{\partial q_{w}}{\partial \varphi}=0$ in the range of $0 \leq \varphi^{\ast} \leq \lambda_{\min}$, 
\begin{eqnarray}
\label{eq62}
\varphi^{\ast}\eq\frac{\left(\kappa-\sqrt{\alpha\kappa(\kappa-1)}\right) \left(\kappa-1-\sqrt{\alpha\kappa(\kappa-1)}\right) }{\left(\kappa-\frac{\alpha}{\alpha-1}\right)\left(\kappa+\frac{1}{\alpha-1}\right)},
\end{eqnarray}
and the maximal investment concentration $q_{w, \max}=q_w(\varphi^*)$ is calculated as follows:
\begin{eqnarray}
\label{eq63}
q_{w,\max}
&=&\frac{\left(\sqrt{\alpha \kappa}+\sqrt{\kappa-1}\right)^2}{\alpha-1}.
\end{eqnarray}
This result also is consistent with results of previous work \cite{11}.

\subsection{Minimal investment concentration with constraints of budget and investment risk}
Finally, we consider the minimal investment concentration with constraints of budget and investment risk as the dual problem. 
The Lagrange function corresponding to the minimization problem of the investment concentration $\L_{D}(\vec{w},h,\varphi)$ is defined as follows:
\begin{eqnarray}
\label{eq77}
\L_{D}(\vec{w},h,\varphi)\eq \H_{D}(\vec{w})+\frac{h}{\varphi} \left(\vec{w}^{\rm T}\vec{e}-N \right)
+\frac{1}{\varphi} \left(\kappa N \varepsilon_{0}-\frac{1}{2}\vec{w}^{{\rm T}}J\vec{w} \right),\nn
{\rm s.t.}&&h \in {\bf R} \nonumber \\
&&\varphi^{-1}\lambda_{i} \leq 1\qquad \forall i=1,2,...,N,
\end{eqnarray}
where the constraint in Eq.~(\ref{eq77}) is a condition to guarantee that the Lagrange function $\L_{D}(\vec{w},h,\varphi)$ is concave with respect to $\vec{w}$ \cite{Boyd,Nocedal,20}.
This constraint can be rewritten as $\varphi \ge \lambda_{\max}=(1+\sqrt{\alpha})^2$ or $\varphi \leq 0$ by using the Mar$\check{{\rm c}}$enko-Pastur law when the number of assets $N$ is large enough.
Since the optimality of $\vec{w}$ and $h$ have already been shown in Eq.~(\ref{eq56}) and (\ref{eq57}), the minimal investment concentration $q_{w,\min}$ is given by the following equation:
\begin{eqnarray}
\label{eq78}
q_{w,\min}&=&\sup_{\varphi \leq 0, \varphi \ge \lambda_{\max}} q_w(\varphi){,}
\end{eqnarray}
where $q_w(\varphi)$ is in Eq.~(\ref{eq42}). 
{Note that this equation is equivalent to the minimization of investment concentration with respect to portfolio $\vec{w}$.}
We can derive $\varphi^{\ast}$ of $\frac{\partial q_{w}(\varphi)}{\partial \varphi}=0$ in the ranges of $\varphi \leq 0$ and $\varphi \ge \lambda_{\max}$ as follows:
\begin{eqnarray}
\label{eq79}
\varphi^{\ast}\eq\frac{\left(\kappa+\sqrt{\alpha\kappa(\kappa-1)}\right) \left(\kappa-1+\sqrt{\alpha\kappa(\kappa-1)}\right) }{\left(\kappa-\frac{\alpha}{\alpha-1}\right)\left(\kappa+\frac{1}{\alpha-1}\right)}, 
\end{eqnarray}
where there exists $\varphi^{\ast}$ such that $\varphi^{\ast} \leq 0$ when  $\kappa < \frac{\alpha}{\alpha-1}$ and $\varphi^{\ast} \ge \lambda_{\max}$ when $\kappa > \frac{\alpha}{\alpha-1}$.
Thus, the minimal investment concentration $q_{w, \min}=q_w(\varphi^*)$ is obtained as follows:
\begin{eqnarray}
\label{eq80}
q_{w,\min}&=&\frac{\left(\sqrt{\alpha \kappa}-\sqrt{\kappa-1}\right)^2}{\alpha-1}.
\end{eqnarray}
This result also is consistent with results of previous work \cite{11}.

From our results for the dual problem, if we assume that the maximal investment concentration $q_{w, \max}$ in Eq.~(\ref{eq63}) is equal to $\tau$, then since $\kappa$ can be represented as $\kappa=\frac{\alpha \tau + \tau -1 - 2\sqrt{\alpha \tau (\tau-1) }}{\alpha-1}$, the investment risk per asset $\varepsilon=\kappa \varepsilon_{0}$ is as follows:
\begin{eqnarray}
\label{eq}
\varepsilon\eq\frac {\alpha \tau+\tau-1-2\sqrt {\alpha \tau \left( \tau -1\right) }}{2},
\end{eqnarray}
which agrees with the minimal investment risk per asset $\varepsilon_{\min}$ in Eq.~(\ref{eq54}).
That is, the primal-dual relationship holds between the minimization of investment risk per asset under constraints of budget and investment concentration and the maximization of investment concentration under constraints of budget and investment risk. 
Similarly, if we also assume that the minimal investment concentration  $q_{w, \min}$ in Eq.~(\ref{eq80}) is consistent with $\tau$, then since $\kappa$ can be represented as $\kappa=\frac{\alpha \tau + \tau -1 + 2\sqrt{\alpha \tau (\tau-1) }}{\alpha-1}$, the investment risk per asset $\varepsilon=\kappa \varepsilon_{0}$ is as follows:
\begin{eqnarray}
\label{eq}
\varepsilon\eq\frac {\alpha \tau+\tau-1+2\sqrt {\alpha \tau \left( \tau -1\right) }}{2},
\end{eqnarray}
which agrees with the maximal investment risk per asset $\varepsilon_{\max}$ in Eq.~(\ref{eq30}).
That is, the primal-dual relationship also holds between the maximization of investment risk under constraints of budget and investment concentration and the minimization of investment concentration under constraints of budget and investment risk.
\section{Numerical experiments}
In the analysis in the previous section, we used the asymptotic eigenvalue distribution based on random matrix theory without a detailed discussion of whether the upper and lower bounds of investment risk per asset $\varepsilon$ and investment concentration $q_{w}$ in actual investment market size can be evaluated or not.
In fact, it has not been confirmed whether the theoretical results (Eqs.~({\ref{eq70}), ({\ref{eq30}), ({\ref{eq63}), and ({\ref{eq80})) are valid or not when the number of assets $N$ is sufficiently large but finite (not the thermodynamical limit of $N$).
In this section, we confirm the consistency of theoretical results using the asymptotic eigenvalue distribution by calculating typical values of upper and lower bounds of investment risk and investment concentration in numerical experiments for the case that $N$ is sufficiently large but finite. In numerical experiments, the return rates of the assets $x_{i \mu}$ are taken to be independently and identically distributed according to the standard normal distribution and $M$ return rate matrices $X^{m}=\left \{\frac{x_{i \mu}^{m}}{\sqrt N} \right \}  \in {\bf R}^{N\times p} (m=1,2,...,M)$ are prepared as sample sets.
Furthermore, the number of assets in a numerical simulation $N$ is set as $N=1000$ and the period ratio $\alpha=p/N$ is set as $\alpha=2$ (also, $M=100$ is assumed). 
We assess the optimal solutions by using the steepest descent method for the Lagrange function defined by each return rate matrix $X^{m}$ and then calculate investment risk $\varepsilon^m$ and investment concentration $q_{w}^m$.
We also evaluate their sample averages,
\begin{eqnarray}
\label{eq}
\varepsilon\eq\frac{1}{100} \sum_{m=1}^{100} \varepsilon^m,\\
\label{eq}
q_{w}\eq\frac{1}{100} \sum_{m=1}^{100} q_{w}^m,
\end{eqnarray}
and compare them with 
the findings derived using our proposed approach.

Firstly, the results of the primal problem are considered, shown in Figs.~\ref{fig1} and \ref{fig2}.
Fig.~\ref{fig1} shows the minimal investment risk per asset $\varepsilon_{\min}$ and Fig.~\ref{fig2} shows the maximal investment risk per asset $\varepsilon_{\max}$.
In both figures, the vertical axis is the investment risk per asset $\varepsilon$ and the horizontal axis is the coefficient of the investment concentration constraint $\tau$. 
The solid lines (red) represent theoretical results and the symbols with error bars (black and gray) represent the results of numerical experiments. 
Using these figures, we confirm that the upper and lower bounds of the investment risk per asset for $N=1000$ assets (finite market size) can be evaluated with the theoretical results, since the results for both cases are equivalent.

Next, the results of the dual problem are considered. Figs.~\ref{fig3} and \ref{fig4}
show respectively the maximal investment concentration $q_{w,\max}$ and the minimal investment concentration $q_{w,\min}$.
In both figures, the vertical axis is investment concentration $\varepsilon$ and the horizontal axis is the risk coefficient $\kappa$.
The solid lines (red) represent theoretical results and the symbols with error bars (black and gray) represent the results of numerical experiments. 
Using these figures, we also confirm that the upper and lower bounds of the investment concentration for $N=1000$ assets (finite market size) can be evaluated with the theoretical results since in both cases the results are consistent.

\begin{figure}[t]
\includegraphics[keepaspectratio,
width=0.9\hsize]{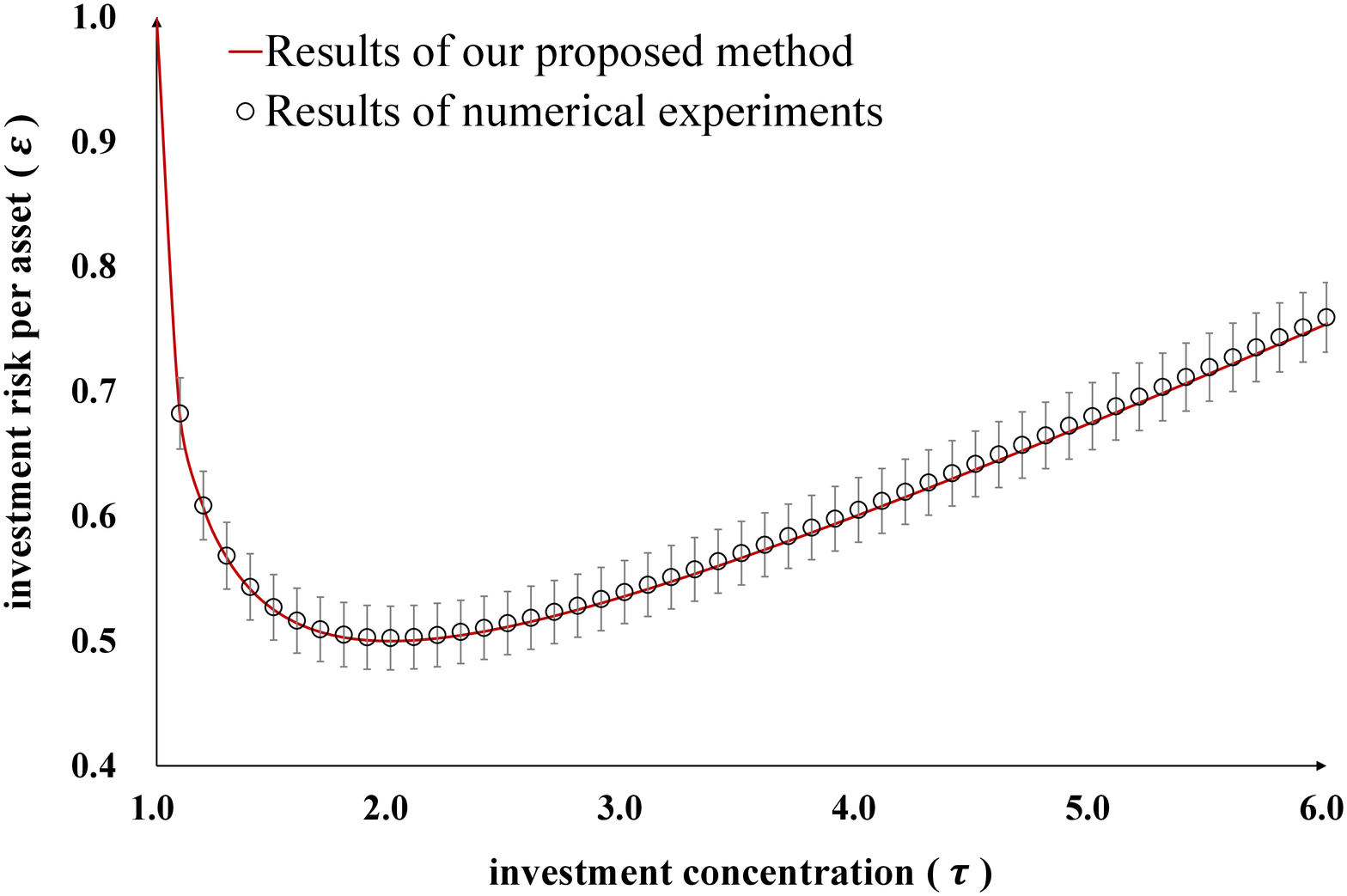}
\caption{Minimal investment risk per asset}
\label{fig1}
\includegraphics[keepaspectratio, width=0.9\hsize]{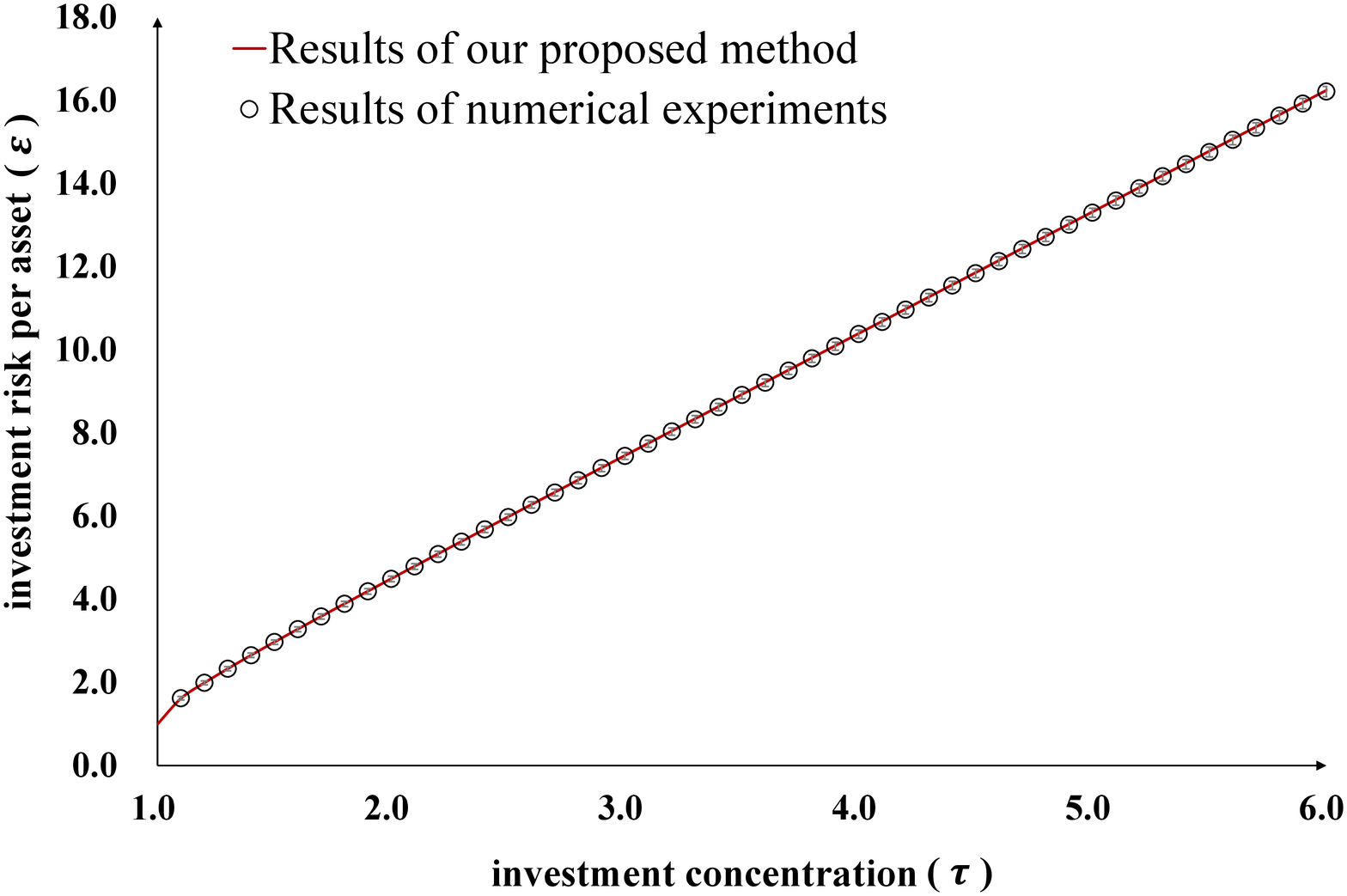}
\caption{Maximal investment risk per asset} 
\label{fig2}
\end{figure}
\begin{figure}[t]
\includegraphics[keepaspectratio, width=0.9\hsize]{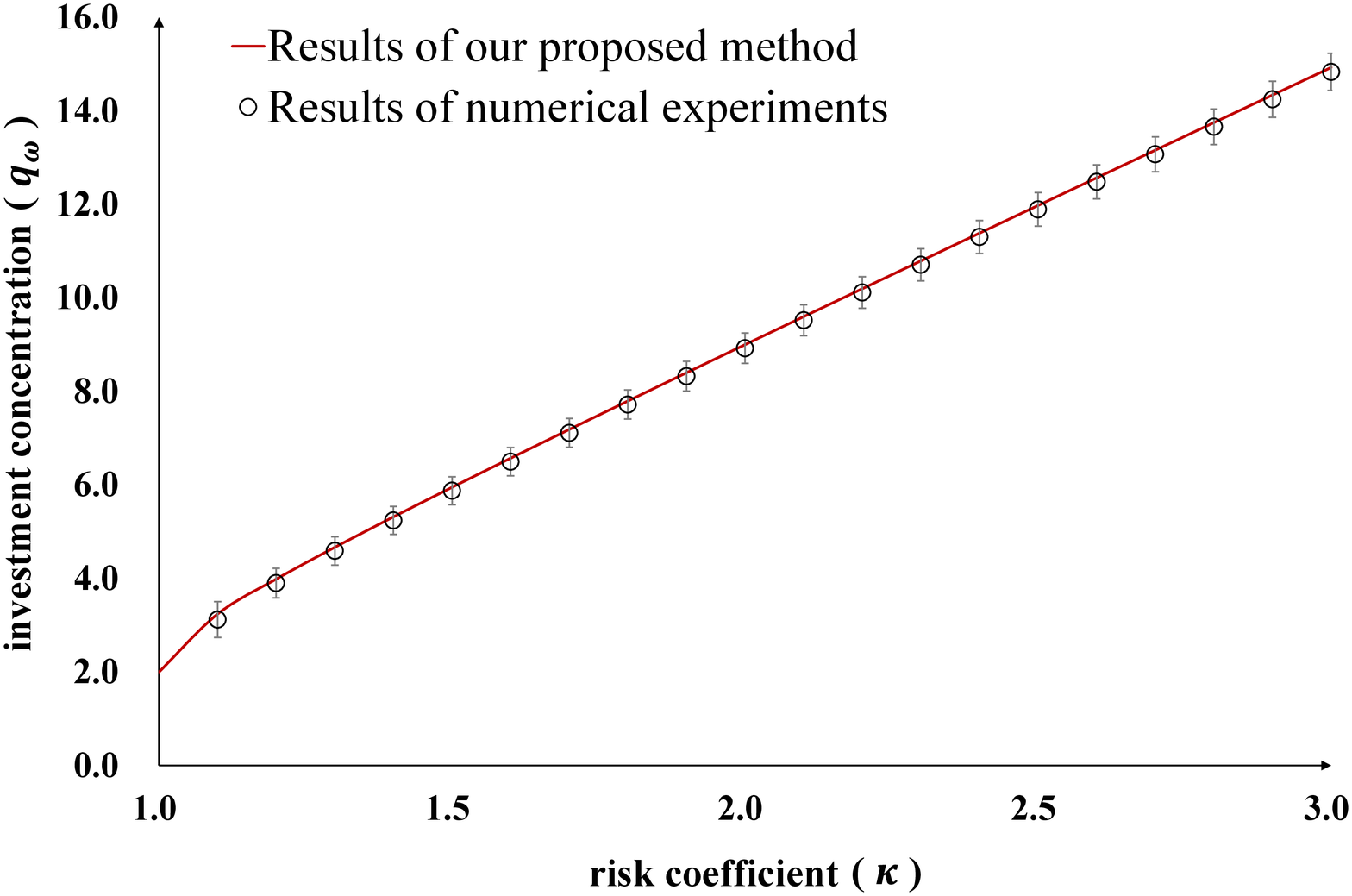}
\caption{Maximal investment concentration} 
\label{fig3}
\includegraphics[keepaspectratio,
width=0.9\hsize]{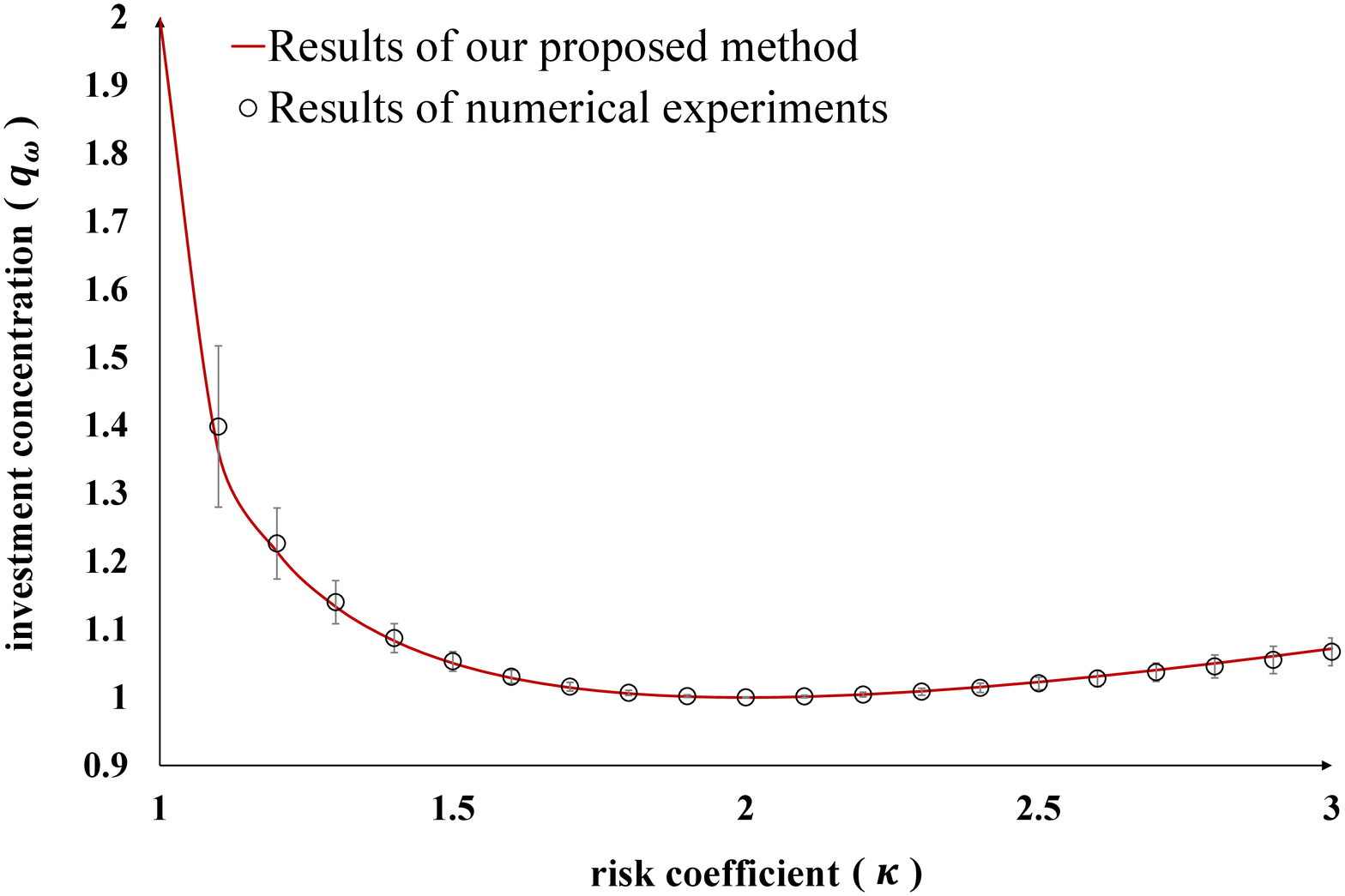}
\caption{Minimal investment concentration} 
\label{fig4}
\end{figure}
\section{Conclusion and future work}
In this paper, 
we revisited the portfolio optimization problem, specifically the minimization/maximization of investment risk under constraints of budget and investment concentration (primal problem) and the maximization/minimization of investment concentration under constraints of budget and investment risk (dual problem).
Both problems had already been analyzed using replica analysis in previous work.
However, since the validity of the analytical continuation of the replica number from integer to real values has not been mathematically guaranteed, we formally reconsidered both problems by using the Lagrange multiplier method and the random matrix approach.
Specifically, we expressed investment risk per asset and investment concentration as functions of Lagrange multipliers and applied Stieltjes transformation of the asymptotic eigenvalue distribution of the Wishart matrix so as to accurately assess the investment risk and investment concentration. 
In this assessment, it was possible to easily find the optimal value of each optimization problem. 
In addition, since the results obtained by replica analysis were consistent with the results derived by our proposed approach for both the primal problem and the dual problem, the validity of replica analysis for this portfolio optimization problem was confirmed. 
Furthermore, from the results of numerical experiments, it was confirmed that the upper and lower bounds of the investment risk and the investment concentration when the number of assets is large enough but finite can be consistent with the theoretical results based on the asymptotic eigenvalue distribution, which demonstrates the effectiveness of the random matrix approach and replica analysis for analyzing the portfolio optimization problem.

We showed in this study that investment risk per asset and investment concentration can be obtained analytically using Stieltjes transformation for the case that the variances of the return rates of assets are identical.
As an extension of this result, we could evaluate easily the primal problem and the dual problem for the case that the variances of the return rates of assets are not identical in future work.
Another issue to consider is that two constraints were used in the present study but it is also necessary to examine whether the random matrix approach is suitable for obtaining the optimal solution of the 
portfolio optimization problem under even more realistic conditions, for instance, 
when short selling regulations or other linear inequality constraints are imposed.
\acknowledgements
The authors are grateful for detailed discussions with H.~Hojo, A.~Seo, M.~Aida, Y.~Kainuma, {S.~Masuda}, and X.~Xiao.
One of the authors (DT) also appreciates T.~Nakamura and T.~Ishikawa for their fruitful comments.
This work was supported in part by Grants-in-Aid Nos.~15K20999, 17K01260, and 17K01249; Research Project of the Institute of Economic Research Foundation at Kyoto University; and Research Project No.~4 of the Kampo Foundation.
\appendix
\section{Stieltjes transformation for Mar$\check{{\bf c}}$enko-Pastur distribution}
In this appendix, we evaluate Eq.~(\ref{eq18}) by using the residue theorem.
If we assume that the modified return rates $x_{i \mu}$ 
are independently and identically distributed with mean $E[x_{i \mu}]=0$ and variance $V[x_{i \mu}]=1$, 
then the empirical eigenvalue distribution of 
Wishart matrix $\rho_N(\lambda)=\frac{1}{N} \sum_{k=1}^N \delta(\lambda - \lambda_k)$ converges to the asymptotic eigenvalue distribution called the Mar$\check{{\rm c}}$enko-Pastur distribution in the limit as $N$ approaches infinity as follows \cite{15}:
\begin{eqnarray}
\label{A1}
\rho(\lambda)\eq[1 - \alpha]^{+} \delta(\lambda) + \frac{\sqrt{ [\lambda_{+} - \lambda]^{+} [\lambda - \lambda_{-} ]^{+}}}{2\pi \lambda}.
\end{eqnarray}
where ${[x]^{+}}=\max(x,0)$ and $\lambda_{\pm}=(1 \pm \sqrt{\alpha})^2$. 
Initially, we calculate $S(\theta)$ in the range $\alpha \ge 1$ by using Eq.~(\ref{A1}).
$S(\theta)$ for $\theta \in {\bf C}$ can be rewritten using $\lambda=1+\alpha+\sqrt{\alpha}\left(\xi+\frac{1}{\xi} \right)$ as follows:
\begin{eqnarray}
\label{A2}
\displaystyle S(\theta)\eq\frac{i}{4\pi}\oint_{|\xi|=1}\frac{(\xi^2-1)^2}{(\xi-\xi_0)(\xi-\xi_1)(\xi-\xi_2)(\xi-\xi_3)(\xi-\xi_4)}d\xi,
\end{eqnarray}
where the poles $\xi_{i}, i=0,1,\cdots,4, $ are as follows: 
\begin{eqnarray}
\label{A3}
\xi_{0}&=&0, \nonumber \\
\xi_{1}&=&-\sqrt{\alpha}, \nonumber \\
\xi_{2}&=&-\frac{1}{\sqrt{\alpha}},  \\
\xi_{3}&=&\frac{-(1+\alpha-\theta) + \sqrt{(1+\alpha-\theta)^2-4\alpha}}{2\sqrt{\alpha}}, \nonumber \\
\xi_{4}&=&\frac{-(1+\alpha-\theta) - \sqrt{(1+\alpha-\theta)^2-4\alpha}}{2\sqrt{\alpha}}. \nonumber \nonumber 
\end{eqnarray}
The residues at the poles, ${\rm Res}[\xi_i]$, $i=0,1,\cdots,4$, are 
\begin{eqnarray}
\label{A4}
{\rm Res}[\xi_0]&=&1, \nonumber \\
{\rm Res}[\xi_1]&=&\frac{\alpha-1}{\theta}, \nonumber \\
{\rm Res}[\xi_2]&=&-\frac{\alpha-1}{\theta}, \\
{\rm Res}[\xi_3]&=&\frac{\sqrt{(1+\alpha-\theta)^2-4\alpha}}{\theta}, \nonumber \\
{\rm Res}[\xi_4]&=&-\frac{\sqrt{(1+\alpha-\theta)^2-4\alpha}}{\theta}.  \nonumber 
\end{eqnarray}
Since $\xi_1\xi_2=1$ and $\xi_3\xi_4=1$ are satisfied, one of $\xi_1$ ($\xi_3$) and $\xi_2$ ($\xi_4$) satisfies $|\xi|<1$ and the other satisfies $|\xi|>1$.
Thus, owing to $|\xi_{1}|>1$ because $\alpha \ge 1$, the combination of poles present inside the unit circle $|\xi|=1$ is (i) $\xi_0, \xi_2, \xi_3$ when $|\xi_3| < 1$ and  (ii) $\xi_0, \xi_2, \xi_4$ when $|\xi_4| < 1$.
Furthermore, $|\xi_3| < 1$ and $|\xi_4| < 1$ can be rewritten as $\lambda_{-} > \theta$ and $\lambda_{+} < \theta$, respectively, for $\theta \in {\bf R}$, so that $S(\theta)$ becomes the real-valued function given by the following equation:
\begin{eqnarray}
\label{A5}
S(\theta)\eq\left\{ \begin{array}{ll} 
\frac{\alpha-1-\theta - \sqrt{\left(1+\alpha-\theta \right)^{2}-4\alpha }}{2\theta} & \lambda_{-} > \theta \\ \\
\frac{\alpha-1-\theta + \sqrt{\left(1+\alpha-\theta \right)^{2}-4\alpha }}{2\theta} & \lambda_{+} < \theta,
\end{array} \right.
\end{eqnarray}
In contrast, the real part of $S(\theta)$ diverges in the range $\lambda_{-} \leq \theta \leq \lambda_{+}$ since the denominator of the integrand can be 0.
As a result, for example, since there does not exist an upper bound of $\f{1}{2}\left(\f{1}{S(\theta)}+\tau\theta\right)$ in the range $\lambda_{-} \leq \theta \leq \lambda_{+}$, we cannot solve Eq.~(\ref{eq20}). That is, the optimization problems derived with the Lagrange function cannot be solved in the range $\lambda_{-} \leq \theta \leq \lambda_{+}$.
For this reason, in this paper, we consider only the optimization problems for $\lambda_{-} > \theta$ and $\lambda_{+} < \theta$, where $S(\theta)$ takes only real values. Furthermore, in the range $0<\alpha<1$, it is also possible to calculate poles and fortunately the same result as Eq.~(\ref{A5}) is obtained.
\section{Replica approach for Stieltjes transformation}
In this appendix, we reexamine $S(\theta)$ from a different direction than in Appendix A.
Initially, we consider a partition function $Z$ to obtain $S(\theta)$ as follows:
\begin{eqnarray}
Z\eq\area \f{d\vec{w}}{(2\pi)^{\f{N}{2}}}e^{-\f{1}{2}\vec{w}^{\rm T}(J-\theta I_N)\vec{w}}.
\end{eqnarray}
Then, the logarithm of this partition function is summarized as follows:
\begin{eqnarray}
\log Z\eq-\f{1}{2}\log\det|J-\theta I_N|{.}
\end{eqnarray}
$S(\theta)$ can be obtained using the following 
identities: 
\begin{eqnarray}
\label{eq-b3}
S(\theta)\eq2\pp{\phi(\theta)}{\theta}, \\
\label{eq-c4}
\phi(\theta)\eq\lim_{N\to\infty}\f{1}{N}E[\log Z],
\end{eqnarray}
where configuration averaging is performed in Eq.~(\ref{eq-c4}) since $S(\theta)$ satisfies the self-averaging property in a similar way to in Sec.~\ref{sec3.1}.
From Eqs.~(\ref{eq-b3}) and (\ref{eq-c4}), it can be seen that 
we need to analytically assess 
$\phi(\theta)$ in order to derive 
$S(\theta)$. 
Therefore, $\phi(\theta)$ is calculated with $E[Z^n]$ for $n\in{\bf Z}$ as follows:
\begin{eqnarray}
\label{eq-c5}
\phi(n,\theta)\eq\lim_{N\to\infty}\f{1}{N}\log E[Z^n]\nn
\eq
\mathop{\rm Extr}_{Q_w,\tilde{Q}_w}
\left\{
-\f{\a}{2}\log\det|I_n+Q_w|
+\f{1}{2}{\rm Tr}Q_w\tilde{Q}_w\right.\nn
&&\left.
-\f{1}{2}\log\det|\tilde{Q}_w-\theta I_n|
\right\},
\end{eqnarray}
where $Q_w=\left\{q_{wab}\right\}$ and $\tilde{Q}_w=\left\{\tilde{q}_{wab}\right\}$ are $n \times n$ symmetric matrices and $I_n$ is the $n \times n$ identity matrix. 
We will also use the notation $\mathop{\rm Extr}_A f(A)$ as the extrema of $f(A)$ with respect to $A$.
Furthermore, extremum conditions with respect to $Q_w, \tilde{Q}_w$ are the following equations:
\begin{eqnarray}
\label{eq-c6}
\tilde{Q}_w\eq\a(I_n+Q_w)^{-1},\\
\label{eq-c7}
Q_w\eq(\tilde{Q}_w-\theta I_n)^{-1}.
\end{eqnarray}
Since Eqs.~(\ref{eq-c6}) and (\ref{eq-c7}) contain only $I_n$ besides terms $Q_w$ and $\tilde{Q}_w$, $Q_w$ and $\tilde{Q}_w$ of the extremum conditions can be written as scalar multiples of $I_n$; that is, $Q_w$ and $\tilde{Q}_w$ can be expressed as $Q_w=\chi_w I_n$ and $\tilde{Q}_w=\tilde{\chi}_wI_n$, respectively.
From this, $\chi_w$ and $\tilde{\chi}_w$ 
are obtained as follows:
\begin{eqnarray}
\label{eq-c8}
\chi_w\eq\f{\a-\theta-1+c\sqrt{(1+\a-\theta)^2-4\a}}{2\theta},\\
\label{eq-c9}
\tilde{\chi}_w\eq\a-1-\theta\chi_w\nn
\eq\f{\a-1+\theta-c\sqrt{(1+\a-\theta)^2-4\a}}{2},
\end{eqnarray}
where $c=\pm1$. Then $\phi(n,\theta)$ is as follows:
\begin{eqnarray}
\phi(n,\theta)\eq-\f{n\a}{2}\log(1+\chi_w)+\f{n}{2}\chi_w\tilde{\chi}_w-\f{n}{2}\log(\tilde{\chi}_w-\theta).
\end{eqnarray}
Since this result satisfies $\phi(n,\theta)=n\phi(1,\theta)$, the following holds:
\begin{eqnarray}
\lim_{N\to\infty}\f{1}{N}\log E[Z^n]\eq
\lim_{N\to\infty}\f{1}{N}\log (E[Z])^n.
\end{eqnarray}
That is, $E[Z^n]\simeq (E[Z])^n$ for sufficiently large $N$.
From this, $E[\log Z]$ can be calculated with the Taylor expansion of the logarithmic function, $\log Z=-\sum_{n=1}^\infty\f{(1-Z)^n}{n}$, as follows:
\begin{eqnarray}
E[\log Z]
\eq-\sum_{n=1}^\infty \f{(1-E[Z])^n}{n}\nn
\eq\log E[Z].
\end{eqnarray}
Then $\phi(\theta)\ (=\phi(1,\theta))$ is given by the following equation:
\begin{eqnarray}
\label{eq-b13}
\phi(\theta) \eq-\f{\a}{2}\log(1+\chi_w)+\f{\chi_w\tilde{\chi}_w}{2}-\f{1}{2}\log(\tilde{\chi}_w-\theta). 
\end{eqnarray}
$\chi_w$ in Eq.~(\ref{eq-c8}) and $\tilde{\chi}_w$ in Eq.~(\ref{eq-c9}) each consist of two solutions, from $c=\pm1$.
However, the relevant solution of each solution is selected by the extremum operator in Eq.~(\ref{eq-c5}).
Thus, $\phi(\theta)$ is as follows:
\begin{eqnarray}
\phi(\theta)\eq
\max(\phi_-(\theta),\phi_+(\theta))
\nn
\eq
\left\{
\begin{array}{ll}
\phi_{-}(\theta)
&\l_->\theta\\
\phi_{+}(\theta)
&\l_+<\theta,
\end{array}
\right.
\end{eqnarray}
where $\phi_{-}(\theta)$ means the $\phi(\theta)$ evaluated with $\chi_w$ and $\tilde{\chi}_w$ in the case of $c =-1$ and $\phi_{+}(\theta)$ means the $\phi(\theta)$  evaluated with $\chi_w$ and $\tilde{\chi}_w$ in the case of $c =+1$.
As a result, from Eqs.~(\ref{eq-b3}) and (\ref{eq-b13}), $S(\theta)$ is given by the following equation:
\begin{eqnarray}
S(\theta)\eq\chi_w\nn
\eq\left\{ \begin{array}{ll} 
\frac{\alpha-1-\theta - \sqrt{\left(1+\alpha-\theta \right)^{2}-4\alpha }}{2\theta} & \lambda_{-} > \theta \\ \\
\frac{\alpha-1-\theta + \sqrt{\left(1+\alpha-\theta \right)^{2}-4\alpha }}{2\theta} & \lambda_{+} < \theta,
\end{array} \right.
\end{eqnarray}
where $\chi_w=\f{1}{\tilde{\chi}_w-\theta}$ is used to derive this equation.
This result is consistent with Eq.~(\ref{A5}), which was derived using the residue theorem.
\section{Minimal investment risk and investment concentration with budget constraint\label{app-c}}
In the main part of this paper, we showed that investment risk per asset $\varepsilon$ and investment concentration $q_{w}$ can be evaluated by using Stieltjes transform $S(\theta)$.
In this appendix, using $S(\theta)$, we derive the minimal investment risk per asset $\varepsilon_{\min}$ and its investment concentration for the portfolio optimization problems with only a budget constraint imposed,
and confirm whether they agree with the results of previous work \cite{9}.
Since only the budget constraint is imposed, the Lagrangian function of this minimization problem is the same as that in Eq.~(\ref{eq12}) with $\theta=0$.
Note that we consider the minimization problem in the range $\alpha > 1$ (when $\a<1$, the optimal solution cannot be uniquely determined).
Thus, the minimal investment risk per asset $\varepsilon_{\min}$ is given by Eq.~(\ref{eq20}) as follows:
\begin{eqnarray}
\label{B1}
\varepsilon_{\min}&=&\lim _{\theta \rightarrow 0 }\frac {1}{2 S \left( \theta \right) }. 
\end{eqnarray}
Its investment concentration is given by substituting 0 into $\theta$ in Eqs.~(\ref{eq13}) and (\ref{eq14}) as follows:
\begin{eqnarray}
\label{B2}
q_{w}\eq\lim _{\theta \rightarrow 0 } \frac{S^{'}(\theta)}{\left( S(\theta)\right)^2}.
\end{eqnarray}
For this calculation, $\lim _{\theta \rightarrow 0 }S(\theta)$ and $\lim _{\theta \rightarrow 0 }S'(\theta)$ are obtained by using $S(\theta)$ in the range $\lambda_{-} > \theta$ as follows:
\begin{eqnarray}
\label{B3}
\lim _{\theta \rightarrow 0 }S(\theta)&=&\frac{1}{\alpha-1}, \\
\lim _{\theta \rightarrow 0 }S'(\theta)&=&\frac{\alpha}{(\alpha-1)^3}.
\end{eqnarray}
Thus, the minimal investment risk per asset $\varepsilon_{\min}$ and its investment concentration $q_{w}$ are as follows:
\begin{eqnarray}
\label{}
\varepsilon_{\min}\eq\frac{\alpha-1}{2}, \\
q_{w}\eq\frac{\alpha}{\alpha-1}.
\end{eqnarray}
We can confirm that these results agree with those of previous work \cite{9}.
In addition, it is possible to obtain the minimal investment risk per asset $\varepsilon_{\min}$ and investment concentration $q_{w}$ for the case that the variances of return rates of the assets are not identical  if we can evaluate the Stieltjes transformation $S(\theta)$ corresponding to this optimization problem.
\section{Minimal investment risk with budget and inequality investment concentration constraints}
In this appendix, we evaluate the minimal investment risk per asset with a budget constraint and an inequality constraint on investment concentration which is an extension of the equality constraint in Eq.~(\ref{eq7}) in the primal problem, as follows:
\begin{eqnarray}
\label{eq32}
\sum_{i=1}^N w_i^2 &\ge& N\tau\qquad\tau \ge 1,
\end{eqnarray}
\begin{eqnarray}
\label{eq33}
\sum_{i=1}^N w_i^2 &\leq& N\tau\qquad\tau \ge 1.
\end{eqnarray}
If we assume that the constraint in Eq.~(\ref{eq32}) is imposed instead of Eq.~(\ref{eq7}), then the minimal investment risk per asset $\varepsilon_{\min}$ is assessed as follows:
\begin{eqnarray}
\label{eq34}
\varepsilon_{\min}&=&\sup_{\theta \leq 0} \ \varepsilon(\theta),
\end{eqnarray}
where Karush-Kuhn-Tucker conditions are used \cite{Boyd,Nocedal,20}.
$\theta^{\ast}=\arg \sup_{\theta \leq 0}\varepsilon(\theta)$ in the range $\theta \leq 0$ is as follows:
\begin{eqnarray}
\label{eq35}
\theta^{\ast} &=&\left\{ \begin{array}{ll} 
1+\alpha - \left(2\tau-1 \right) \sqrt{\frac{\alpha}{\tau(\tau-1)}} & \frac{\tau-1}{\tau}<\alpha < \frac{\tau}{\tau-1} \\ \\ \\
 0 &  {\rm otherwise}. 
\end{array} \right. 
\end{eqnarray}
Thus, the minimal investment risk per asset corresponding to Eq.~(\ref{eq35}) is as follows:
\begin{eqnarray}
\label{eq36}
\varepsilon_{\min}&=&\left\{ \begin{array}{ll} 
0  & \alpha \leq \frac{\tau-1}{\tau} \\ \\
\frac {\alpha \tau + \tau -1 -2\sqrt {\alpha \tau \left( \tau -1\right) }}{2}  &  \frac{\tau-1}{\tau}<\alpha < \frac{\tau}{\tau-1}  \\ \\
\frac {\alpha-1}{2}  &  \alpha \ge \frac{\tau}{\tau-1}.
\end{array} \right.
\end{eqnarray}
In contrast, if we apply the constraint in Eq.~(\ref{eq33}) instead of Eq.~(\ref{eq7}), then the minimal investment risk per asset $\varepsilon_{\min}$ is given by the following equation:
\begin{eqnarray}
\label{eq40}
\varepsilon_{\min}&=&\sup_{0 \leq \theta \leq \lambda_{\min}} \varepsilon(\theta),
\end{eqnarray}
where Karush-Kuhn-Tucker conditions are also used \cite{Boyd,Nocedal,20}.
$\theta^{\ast}=\arg \sup_{0 \leq \theta \leq \lambda_{\min}}\varepsilon(\theta)$ in the range $0 \leq \theta \leq \lambda_{\min}$ is as follows:
\begin{eqnarray}
\label{eq38}
\theta^{\ast}&=&\left\{ \begin{array}{ll} 
1+\alpha - \left(2\tau-1 \right) \sqrt{\frac{\alpha}{\tau(\tau-1)}} & \alpha>\frac{\tau}{\tau-1}   \\ \\
 0 & {\rm otherwise}. 
\end{array} \right. 
\end{eqnarray}
Thus, the minimal investment risk per asset corresponding to Eq.~(\ref{eq38}) is as follows:
\begin{eqnarray}
\label{eq39}
\varepsilon_{\min}&=&\left\{ \begin{array}{ll} 
0 & 0 < \alpha < 1 \\ \\
\frac{\alpha-1}{2} & 1\leq \alpha \leq \frac{\tau}{\tau-1} \\ \\
\frac {\alpha \tau + \tau -1 -2\sqrt {\alpha \tau \left( \tau -1\right) }}{2} &  \alpha>\frac{\tau}{\tau-1}. 
\end{array} \right.
\end{eqnarray}
This result also is consistent with that of previous work \cite{10}. 
 
\end{document}